\documentclass[12pt,preprint]{emulateapj}

\usepackage{comment}
\usepackage{xspace}

\shorttitle{Winds and Jets in 4U 1630--47}
\shortauthors{Neilsen et al.}

\begin{document}

\title{A Link Between X-ray Emission Lines and Radio Jets in 4U 1630--47?}

\author{Joseph Neilsen\altaffilmark{1,2,9}, Micka\"el Coriat\altaffilmark{3}, Rob Fender\altaffilmark{4}, Julia C. Lee\altaffilmark{5}, Gabriele Ponti\altaffilmark{6}, Anastasios K. Tzioumis\altaffilmark{7}, Philip G. Edwards\altaffilmark{7}, Jess W. Broderick\altaffilmark{8}}
\altaffiltext{1}{Department of Astronomy, Boston University, Boston, MA 02215, USA; neilsenj@bu.edu}
\altaffiltext{2}{MIT Kavli Institute for Astrophysics and Space Research, Cambridge, MA 02138, USA}
\altaffiltext{3}{Department of Astronomy, University of Cape Town, Private Bag X3, Rondebosch 7701, South Africa}
\altaffiltext{4}{Department of Physics, Brasenose College, University of Oxford, Oxford, OX1 3RH, UK}
\altaffiltext{5}{Harvard-Smithsonian Center for Astrophysics, Cambridge, MA 02138}
\altaffiltext{6}{Max Planck Institute fur Extraterrestriche Physik, 85748, Garching, Germany}
\altaffiltext{7}{CSIRO Astronomy and Space Science, Australia Telescope National Facility, PO Box 76, Epping, NSW 1710, Australia}
\altaffiltext{8}{School of Physics and Astronomy, University of Southampton, Highfield, Southampton, SO17 1BJ, UK}
\altaffiltext{9}{Einstein Fellow}

\begin{abstract}
Recently, \citeauthor{DiazTrigo13} reported an \textit{XMM-Newton} detection of relativistically Doppler-shifted emission lines associated with steep-spectrum radio emission in the stellar-mass black hole candidate 4U 1630--47 during its 2012 outburst. They interpreted these lines as indicative of a baryonic jet launched by the accretion disk.  Here we present a search for the same lines earlier in the same outburst using high-resolution X-ray spectra from the \textit{Chandra} High-Energy Transmission Grating Spectrometer. While our observations (eight months prior to the \textit{XMM-Newton} campaign) also coincide with detections of steep spectrum radio emission by the Australia Telescope Compact Array, we find no evidence for any relativistic X-ray emission lines. Indeed, despite $\sim5\times$ brighter radio emission, our \textit{Chandra} spectra allow us to place an upper limit on the flux in the blueshifted Fe\,{\sc xxvi} line that is $\gtrsim20\times$ weaker than the line observed by \citeauthor{DiazTrigo13} We explore several scenarios that could explain our differing results, including variations in the geometry of the jet or a mass-loading process or jet baryon content that evolves with the accretion state of the black hole. We also consider the possibility that the radio emission arises in an interaction between a jet and the nearby ISM, in which case the X-ray emission lines might be unrelated to the radio emission. 
\end{abstract}
                 

\section{INTRODUCTION}
\label{sec:intro}
\defcitealias{DiazTrigo13}{DT13}

Collimated jets are ubiquitous in accreting systems across the mass scale, from protostars (e.g., \citealt{CarrascoGonzalez10}) to supermassive black holes in AGN at the centers of galaxy clusters (e.g., \citealt{McNamara00,Fabian00}). But the omnipresence of jets in astrophysical systems raises important questions. Are all these jets produced by the same processes? Since one of the canonical jet formation mechanisms (the Blandford-Znajek, or BZ, mechanism; \citealt{BlandfordZnajek,Tchekhovskoy11}), involves the rotation power of a black hole, it certainly cannot operate in all systems. An alternative is the Blandford-Payne mechanism (BP; \citealt{BlandfordPayne}), in which the requisite rotation power comes from a rotating accretion disk. As a consequence, one expects BP jets to be baryonic, since they are magnetocentrifugal outflows from a gaseous reservoir. BZ jets, on the other hand, might be electromagnetically dominated.

In this context, it is particularly interesting that accreting stellar-mass black holes appear to produce two different kinds of jets (steady and transient; for reviews see \citealt{FBG04,Fender09}). Observationally, it remains unclear whether these two types of jets correspond to the two mechanisms described above; a suggested association between the transient jets and the BZ mechanism has lately been the subject of significant debate (\citealt{Fender10,Narayan12,Steiner13,Russell13,King13b}). The formation of steady jets is no less significant, as it has been suggested (\citealt{NL09}; see also \citealt{M08,N11a,N12b,King13} and references therein) that ionized accretion disk winds may be able to quench steady jets. Different processes may launch astrophysical jets at different scales or under different conditions, but because a jet's energy requirements are sensitive to its baryon content (see, e.g., \citealt{F99,Gallo05,Heinz06,Punsly13}), these considerations have broad and concrete implications for radio-mode feedback from accreting systems. 

\begin{figure*}
\centerline{\includegraphics[width=\textwidth]{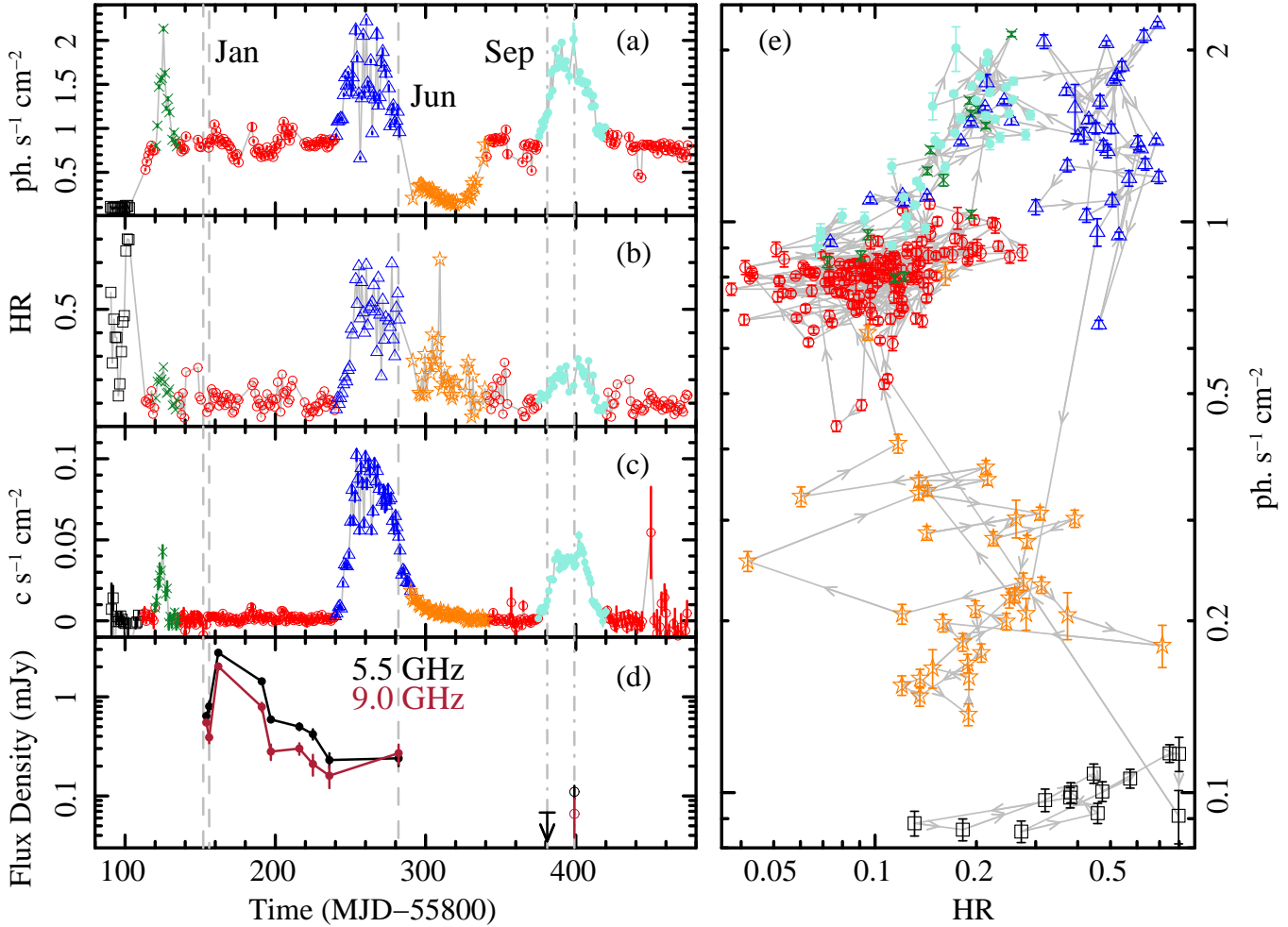}}
\caption{Monitoring of the 2012 outburst of 4U 1630--47. (a) 2--20 keV MAXI light curve. (b) MAXI hardness ratio (10--20/2--4 keV). (c) 15--50 keV \textit{Swift}/BAT light curve. (d) 5.5 GHz and 9 GHz radio fluxes as measured by ATCA by us and \citetalias{DiazTrigo13}. Dashed and dash-dotted vertical lines indicate the times of the \textit{Chandra} and \textit{XMM} observations, respectively. (e) MAXI hardness-intensity diagram. The colors/symbols are rough indicators of the accretion state.\label{fig:lc}}
\end{figure*}

Recently, \citeauthor{DiazTrigo13} (\citeyear{DiazTrigo13}; hereafter \citetalias{DiazTrigo13}) reported a detection of relativistically-blueshifted emission lines associated with radio emission in the stellar-mass black hole candidate 4U 1630--47, a well-known recurrent transient  (\citealt{Jones76,Kuulkers97b,Tomsick05}) that went into outburst in December 2011 (\citealt{N13a}). \citetalias{DiazTrigo13} reported on two observations of the source with \textit{XMM-Newton} and the Australia Telescope Compact Array (ATCA) in 2012 September. In their first observation (September 10-12), they found a featureless X-ray spectrum and no radio emission. In their second observation (September 28-29), however, they discovered optically thin radio emission and X-ray emission lines at energies that do not correspond to strong lines from abundant elements. They argued that the emission lines originated in a relativistic jet with a speed of $\sim0.66c,$ where $c$ is the speed of light. In their interpretation, the lines corresponded to red- and blueshifted Fe\,{\sc xxvi} Ly$\alpha$ and blueshifted Ni\,{\sc xxvii} He$\alpha$.

In this paper, we report on ATCA and \textit{Chandra} High-Energy Transmission Grating Spectrometer (HETGS; \citealt{C05}) observations of the same outburst of 4U 1630--47. Although we detect radio emission at levels brighter than those reported in \citetalias{DiazTrigo13}, we do not detect the X-ray emission lines seen later in the outburst. Indeed, our \textit{Chandra} spectra place tight upper limits on such features, and we can conclusively say that radio emission is not universally associated with relativistically Doppler-shifted emission lines in 4U 1630--47. While our results do not and cannot directly address the argument for the presence of baryons in the 2012 September jet in 4U 1630--47, we highlight the necessity of a complex relationship between X-ray and radio emission, considering not only the possibility of undetected baryons in the January and June jets or baryonic jets confined to a particular accretion state, but also an alternative scenario in which the radio emission arises in a jet-ISM interaction and could be decoupled from the X-ray emission.

\section{OBSERVATIONS AND DATA REDUCTION}
\label{sec:obs}
\subsection{\textit{Chandra} HETGS}
\label{sec:chandra}
After the peak of the outburst around 2012 January 1, we triggered our \textit{Chandra} HETGS campaign, observing the source for $\sim29$ ks four times: January 17 (4:23:43 UT), January 20 (23:43:52 UT), January 26 (13:00:35 UT), and January 30 (8:48:41 UT). The data were taken in Graded mode with a grey filter over the zeroth order to reduce the risk of telemetry saturation from the bright source. We also obtained a 19 ks HETGS DDT in Continuous-Clocking (CC) mode (2012 June 3 22:19:16 UT) when 4U 1630--47 entered an active state. We show the MAXI/\textit{Swift} outburst light curves and MAXI hardness-intensity diagram in Figure \ref{fig:lc}. We reduce the \textit{Chandra} data with standard {\sc ciao} data reduction tools to create first-order High-Energy Grating (HEG) spectra and response files. Performing all our analysis within the Interactive Spectral Interpretation System (ISIS; \citealt{HD00}), we restrict our attention to the 1.7--9 keV band. 

\begin{figure*}
\centerline{\hfill\includegraphics[height=3.75in]{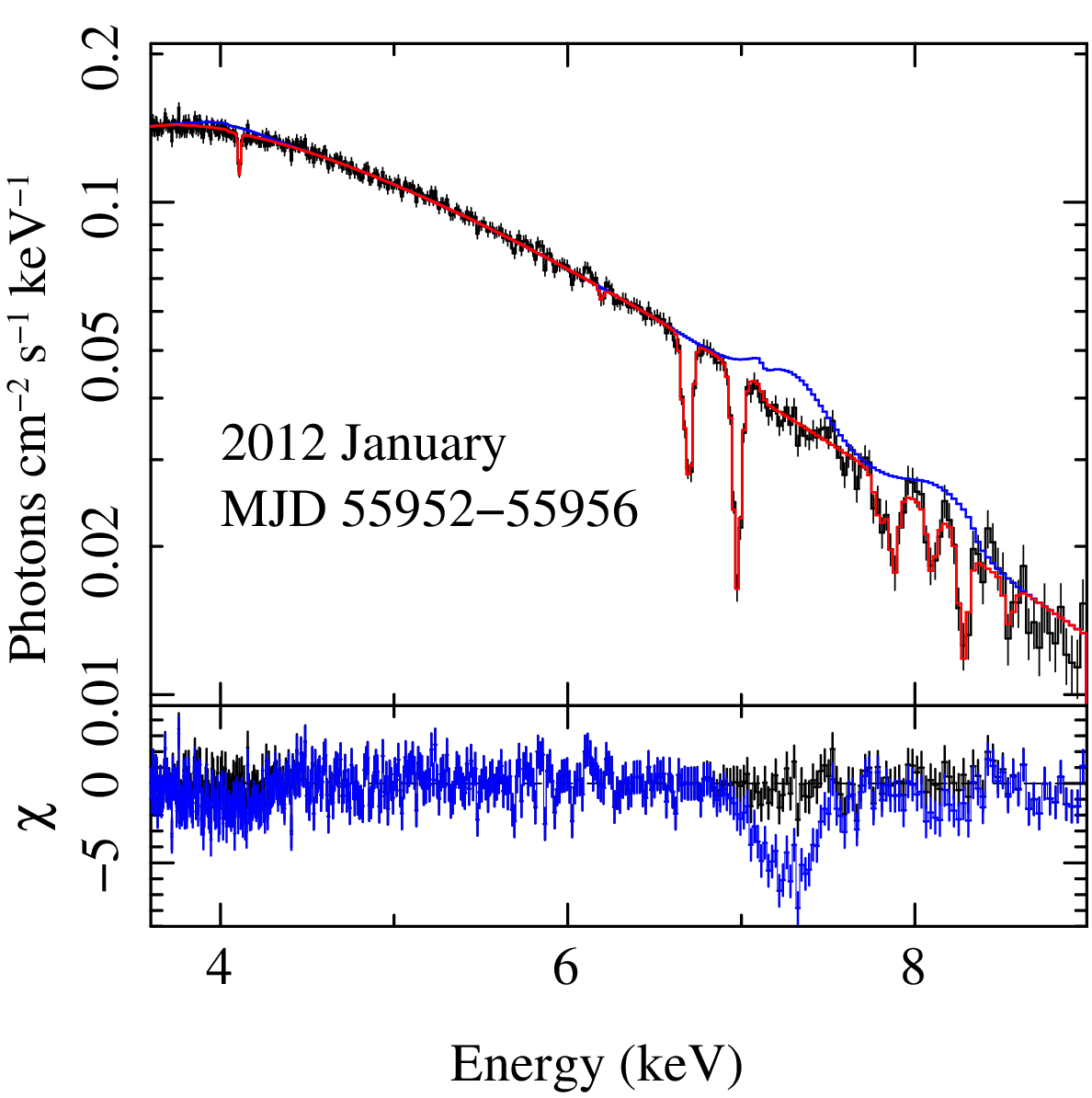}\hfill\includegraphics[height=3.75in,trim=25mm 0mm 0mm 0mm,clip=true]{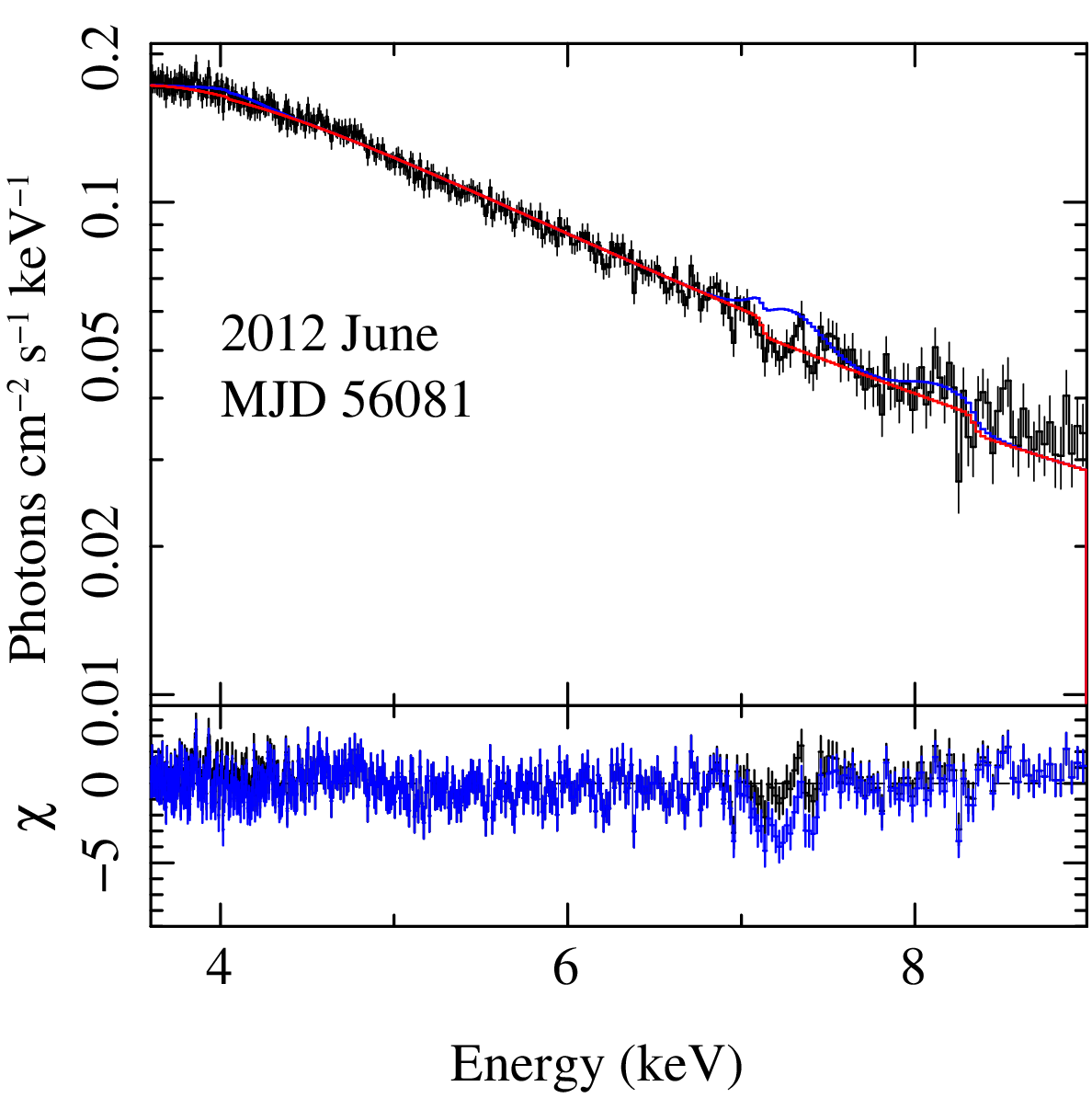}\hfill}
\caption{\textit{Chandra} HETGS spectrum of 4U 1630--47 (black) from January (left) and June (right). In red, we show our best fit models. In blue, we plot the sum of the continuum emission and the relativistic lines seen by \citetalias{DiazTrigo13}. In the bottom panels are the residuals relative to our best fit; the non-detection of the relativistic lines is highly significant. See text for details.\label{fig:spec}}
\end{figure*}

\subsection{ATCA}
\label{sec:atca}
We also triggered an ATCA monitoring campaign at 5.5 and 9 GHz, observing the source at 10 epochs between 2012 January 28 and June 4. The observations were conducted with the Compact Array Broadband Backend (CABB; \citealt{Wilson11}) with the array in a number of configurations ranging from H168 (compact) to 6A (extended). Each frequency band was composed of 2048 1-MHz channels. We used PKS1934--638 for absolute flux and bandpass calibration, and J1600--48 to calibrate the antenna gains as a function of time. Flagging, calibration and imaging were carried out with the Multichannel Image Reconstruction, Image Analysis and Display (MIRIAD) software (\citealt{Sault95}). Note that the visibilities from the shortest baselines were excluded from the imaging step to avoid flux contamination from an extended source very close to 4U 1630--47. 

We measured flux densities by fitting a point source in the image plane, and we achieve a significant detection of 4U 1630--47 at 5.5 and 9 GHz in every observation with useful data. The resulting ATCA light curves are plotted in the bottom panel of Figure \ref{fig:lc} and appear to show an optically thin flare that lasts $\sim50$ days and then fades slowly as the spectrum flattens.

\section{Results}
\label{sec:spec}

During their second observation of 4U 1630--47 (MJD 56181), \citetalias{DiazTrigo13} reported radio fluxes of $110\pm17$ and $66\pm28$ $\mu$Jy at 5.5 GHz and 9 GHz, respectively. In order to perform the most robust comparison, we focus here on the \textit{Chandra} observations with contemporaneous ATCA detections of optically thin radio emission (i.e., January 26 and 30, MJD 55952/55956). The ATCA pointings on January 28 and 30 have 5.5 GHz/9 GHz fluxes of $640\pm30~\mu$Jy/$550\pm30~\mu$Jy and $800\pm100~\mu$Jy/$390\pm50~\mu$Jy, respectively. For completeness, we also consider our June 3-4 (MJD 56081) \textit{Chandra}/ATCA campaign (fluxes of $240\pm40~\mu$Jy and $270\pm60~\mu$Jy, respectively).

\subsection{2012 January}

The most notable feature of our January \textit{Chandra} spectra is a series of deep, blueshifted ($v\sim200-500$ km s$^{-1}$), narrow absorption lines indicative of a highly-ionized accretion disk wind (including Ca\,{\sc xx}, Fe\,{\sc xxv} He\,$\alpha/\beta/\gamma/\delta$, Fe\,{\sc xxvi} Ly\,$\alpha/\beta$, Ni\,{\sc xxvii} He\,$\alpha$, and Ni\,{\sc xxviii} Ly\,$\alpha$; see Figure \ref{fig:spec}). Its presence is consistent with the picture put forward in \citeauthor{Ponti12} (\citeyear{Ponti12}; see also \citealt{N12b,N13a}). \textit{NuSTAR} also detected a wind a month later (\citealt{King14}). For our purposes, the wind is a confounding factor in the search for the emission lines detected by \citetalias{DiazTrigo13}; we will discuss its properties in detail in a forthcoming paper (Neilsen et al., in preparation). Here, we focus on the X-ray continuum and a search for Doppler-shifted emission lines.

We model the combined spectrum of the \textit{Chandra} observations as a pure disk blackbody ({\tt ezdiskbb}; \citealt{Zimmerman05}) modified by interstellar absorption ({\tt TBnew}, using \citealt{Wilms00} abundances and \citealt{Verner96} cross-sections), dust scattering ({\tt dustscat}; \citealt{Baganoff03}), and 17 narrow absorption lines. We tie the dust optical depth $\tau$ to $N_{\rm H}$ via the relationship $\tau=0.324\times10^{-22}$ cm$^{2}~N_{\rm H}$ (see \citealt{Nowak12} for more details on this scaling). We also allow the interstellar abundances of Si, S, and Ni to vary, and we use the model {\tt simple\_gpile2} to account for photon pileup (\citealt{Davis01}). All errors quoted below are 90\% confidence limits for a single parameter.

\begin{deluxetable*}{cccc@{\extracolsep{3mm}}c@{\extracolsep{1mm}}cccc}
\tabletypesize{\scriptsize}
\tablecaption{Doppler-Shifted Emission Lines in 4U 1630--47
\label{tbl:lines}}
\tablewidth{0.5\textwidth}
\tablehead{
\colhead{}  &
\colhead{}  &
\colhead{Jan.}  & 
\colhead{} & 
\colhead{}  &
\colhead{Jun.}  & 
\colhead{} & 
\colhead{} & 
\colhead{} \\
\cline{2-4}\cline{5-7}\\[-2mm]
\colhead{}  &
\colhead{}  &
\colhead{$E_{\rm obs}$}  & 
\colhead{$W_0$} & 
\colhead{}  &
\colhead{$E_{\rm obs}$}  & 
\colhead{$W_0$} & 
\colhead{} & 
\colhead{$W_{\rm DT13}$} \\
\colhead{Line}  &
\colhead{$F_{8}$}  &
\colhead{(keV)}  &
\colhead{(eV)} &
\colhead{$F_{8}$}  &
\colhead{(keV)}  &
\colhead{(eV)} &
\colhead{$F_{\rm DT13}$} & 
\colhead{(eV)}
}
\startdata
\vspace{1mm}
Fe$_{\rm red}$ & $<0.3$ & $4.04$\tablenotemark{a}  & $<2$ &$4\pm1$\tablenotemark{b}&\hspace{1mm}$3.94^{+0.06}_{-0.0p}$\tablenotemark{a}&\hspace{0mm}$18\pm5$\tablenotemark{b}& $3.0\pm1.4$ & $8\pm4$ \\
\vspace{1mm}
Fe$_{\rm blue}$ & $<0.3$ & 7.28\tablenotemark{a} & $<6$ &$<1.3$&7.28\tablenotemark{a}&$<18$& $6.2\pm1$ & $37\pm9$ \\
\vspace{1mm}
Ni$_{\rm blue}$ & $<10$ & 8.14\tablenotemark{a} & $<270$ &$<4.3$&8.14\tablenotemark{a}&$<70$& $2.5\pm1.0$ & $18\pm10$
\enddata
\tablecomments{Constraints on Doppler-shifted emission lines in HETGS spectra of 4U 1630--47. $F_8$ is the emission line flux in units of $10^{-8}$ erg s$^{-1}$ cm$^{-2}$, $E_{\rm obs}$ is the line energy, $W_0$ is the equivalent width, $F_{\rm DT13}$ is the flux measured by \citetalias{DiazTrigo13}, and $W_{\rm DT13}$ is their equivalent width. Errors are 90\% confidence limits for a single parameter. $p$ indicates a pegged limit.}
\tablenotetext{a}{Unconstrained or poorly constrained by the HETGS data. Required to be within the 90\% confidence limits of \citetalias{DiazTrigo13}.}
\tablenotetext{b}{Attributed to calibration effects in CC mode.}
\end{deluxetable*}

Our best fit column density, $N_{\rm H}=9.4_{-1.1}^{+0.5}\times10^{22}$ cm$^{-2},$ is consistent with the value reported by \citetalias{DiazTrigo13}, but the disk in our observations is both cooler and fainter than in \citetalias{DiazTrigo13}'s observations, even when accounting for the differences in our disk model (\citealt{Zimmerman05}). For reference, in their second observation \citetalias{DiazTrigo13} measured a disk temperature of $1.77\pm0.03$ keV and a 2--10 keV flux of $5.2\times10^{-8}$ erg s$^{-1}$ cm$^{-2}.$ In contrast, in January we measure a disk temperature of $T_{\rm max}=1.33\pm0.01$ keV and an unabsorbed 2--9 keV flux of $1.1\times10^{-8}$ erg s$^{-1}$ cm$^{-2}$. For a distance of 10 kpc, this flux corresponds to an unabsorbed luminosity of $\sim1.3\times10^{38}$ erg s$^{-1}.$ Our best fit is shown in red in the left panel of Figure \ref{fig:spec}. 

To search for relativistic lines in our HETGS spectra, we add three Gaussian emission lines to our fit model. We constrain the line energies and widths to be within the 90\% confidence limits reported by \citetalias{DiazTrigo13}: the energies are $4.04\pm0.1$ keV, $7.28\pm0.04$ keV, and $8.14\pm0.16$ keV, and the line width is $0.17\pm0.05$ keV. In Figure \ref{fig:spec}, we show these emission lines at the fluxes reported by \citetalias{DiazTrigo13} (blue curve). In our analysis we allow these fluxes to vary, but none are significant at the $3\sigma$ level. Indeed, we find that relative to the second observation of \citetalias{DiazTrigo13}, the flux in the redshifted and blueshifted Fe\,{\sc xxvi} lines is reduced by factors of $\gtrsim10$ and $\gtrsim20,$ respectively. The blueshifted Ni\,{\sc xxvii} line is poorly constrained because of absorption lines in the HETGS spectrum (Figure \ref{fig:spec}, left).

\setcounter{footnote}{0}
\subsection{2012 June}

In contrast to the pure disk continuum in January, we find a significant power law component in June, which we model using {\tt simpl$\otimes$ezdiskbb} (\citealt{Steiner09b}). The asymptotic photon index is $\Gamma=2.9^{+0.2}_{-0.3}$, but the fraction of scattered disk photons is unconstrained because of the low disk temperature ($T_{\rm max}<0.3$ keV). The unabsorbed 2--9 keV flux is $1.4\times10^{-8}$ erg s$^{-1}$ cm$^{-2}.$ We search for relativistic lines using the same procedure as above. Again we find no statistically-significant blueshifted emission\footnote{There is a small residual near 7.3 keV (Figure \ref{fig:spec}), but an F-test indicates that it is not significant at the 90\% confidence level.} (Table \ref{tbl:lines}). There is a feature detected near $\sim4$ keV, but with significant calibration uncertainties in CC mode associated with high column densities and bright/hard spectra, and no blueshifted counterpart, we believe it may be unphysical and does not support a June baryonic jet. We also note that the associated radio emission is optically thick, so the jet physics likely differs somewhat from that in \citetalias{DiazTrigo13}.

\section{DISCUSSION}
\label{sec:discuss}

In their second \textit{XMM-Newton}/ATCA observation of 4U 1630--47 in 2012 September, \citetalias{DiazTrigo13} discovered relativistically Doppler-shifted Fe and Ni emission lines whose appearance coincided with steep-spectrum radio emission. But despite the $\sim5\times$ brighter steep-spectrum radio emission during our ATCA observations in 2012 January, we found no evidence for the emission lines in our contemporaneous \textit{Chandra} observations. With a tight upper limit on the blueshifted Fe\,{\sc xxvi} emission line flux (it must have been $\gtrsim20\times$ fainter than reported by \citetalias{DiazTrigo13} eight months later), what can be said about the evolution of the baryon content of the jet in 4U 1630--47? What is the significance of \textit{Chandra}'s non-detection of blueshifted emission lines? 

Here we will focus on our January campaign, since it provides the strongest constraints. We can explain the absence of the lines in light of two scenarios: (1) there were relativistic baryons in the jet at the time of our campaign, but their X-ray emission was negligible, or (2) there were no relativistic baryons in the jet. Of course there is middle ground, but these scenarios span the plausible range of explanations. Below, we consider them in turn. In Section \ref{sec:dt13} we also discuss the association between the radio emission and relativistic lines seen by \citetalias{DiazTrigo13}.\vspace{-1mm}

\subsection{2012 January: Baryons but No Emission Lines}

First, we suppose that despite the non-detection of baryonic emission during our \textit{Chandra}/ATCA campaign, the jet had the same baryon content as in the \textit{XMM}/ATCA observations. Specifically, we assume the jet emission mechanisms were the same (but see Section \ref{sec:dt13}), and we consider either the jet baryon number density $n_b$ or the total number of jet baryons $N_b$ to be fixed. These considerations are reasonable in context because \citetalias{DiazTrigo13} modeled the baryonic emission in their \textit{XMM} spectrum as a 21 keV thermal plasma, for which the model normalization $K$ is proportional to $\int n_e n_{\rm H} dV\sim n_b N_b;$ $n_e$ is the electron density, $n_{\rm H}$ is the density of hydrogen, and the integral is over the emitting volume $dV$.

To suppress the plasma emission by $\sim20,$ we can therefore invoke a smaller baryon density or a reduced number of baryons in the jet. Keeping $N_b$ fixed and reducing $n_b$ would require a $\sim20\times$ larger jet, while keeping $n_b$ fixed and reducing the total number of baryons would require a $\sim20\times$ smaller jet. Physically, these scenarios could represent a jet at a different stage of formation. In either case ($n_b$ or $N_b$ fixed), some fine tuning of the plasma parameters would also be needed, since the optically thin synchrotron emission from the jet is sensitive to both the volume and the electron density in the emitting region.

Alternatively, we might explain the non-detection of the iron emission line by invoking a variation in the plasma temperature that significantly reduces the abundance of Fe\,{\sc xxvi} and its emissivity. However, we can effectively rule out such a temperature variation by querying the APEC database that underlies the {\tt bvapec} model used by \citetalias{DiazTrigo13}: the emissivity of Fe\,{\sc xxvi} drops to 5\% of its value at temperatures (1) below $\sim3$ keV, where emission from Fe\,{\sc xxv} would have been even more apparent, and (2) well above 100 keV, which would be difficult to reconcile with our cooler accretion disk and no apparent power law component in the X-ray spectrum. As noted in \citetalias{DiazTrigo13}, extreme line broadening could also effectively hide such lines, but the iron line would need to be nearly an order of magnitude broader to be undetectable with \textit{Chandra}, and it is difficult to explain such a large change in the expansion speed of the jet.

\subsection{2012 January: No Baryons}
\label{sec:state}
Perhaps the simplest interpretation is the one that takes our data at face value: we did not observe relativistic iron emission lines with \textit{Chandra} because there were no baryons in the 2012 January jet. In this interpretation, the question is not why we did not observe the relativistic emission lines, but why \citetalias{DiazTrigo13} did. What changed in the accretion flow between our campaign and theirs (and, for that matter, between their two reported \textit{XMM} observations)?

The most significant difference in the observations appears to be the accretion state. As noted above, the X-ray luminosity of 4U 1630--47 in the second XMM observation was $\sim3-5\times$ higher than in January and June. Notably, that luminosity corresponds to a significant fraction ($\sim50\%$) of the Eddington luminosity for a 10 $M_{\odot}$ black hole at a distance of 10 kpc, and $\gtrsim50\%$ of the emission was in a hard power law. These facts raise the possibility that the stronger, brighter non-thermal component, the very high accretion power, or the extreme luminosity of the source may have played a role in lifting and/or helping to accelerate heavy particles.

\subsection{2012 September: Jet-ISM Interaction?}
\label{sec:dt13}
It is also worth considering whether or not the relativistic emission lines are indicative of baryonic matter in a relativistic jet, or if they can truly be associated with the radio emission at all. The primary reasons for this are that radio jets have been somewhat elusive in 4U 1630--47 (\citealt{Tomsick05}), and our ATCA detections in January are atypical for steady jets from stellar-mass black holes. As mentioned in Section \ref{sec:atca}, our radio light curve looks like a faint, optically thin flare that lasts many weeks. Since our initial ATCA detections of very faint optically thin emission occur $\sim50$ days after a major hard-soft state transition (Fig.\ \ref{fig:lc}; MJD $\sim55900-55910$), and in what appears to be a standard disk-dominated state, we are inclined to interpret them as indicative of a jet-ISM interaction far from the black hole. See \citet{Fender09} and references therein for a thorough discussion of similar detections in spectrally soft states.

Given the similar optically-thin spectrum, could the 2012 September 28 radio emission also originate in a jet-ISM interaction? Inspection of the light curves and hardness-intensity diagram in Figure \ref{fig:lc} reveals that between June and September there was a transition from a spectrally hard state to a softer state, although the date is difficult to identify precisely. Such transitions are usually associated with a transient ejection, and the optically thick radio emission in June might indicate the start of radio flaring periods typical of these transitions. If the associated ejection reached the ISM $2-4$ months later, it could potentially explain the radio emission in the second \textit{XMM}/ATCA observation. If that is the case, then the relativistic lines might not have been associated with the radio source at all, i.e., the radio emission could have been produced in the nearby ISM, while the X-ray lines were local to 4U 1630--47. For example, at a temperature of 21 keV, a relativistic wind would appear in emission in front of the accretion disk, although some obscuration, absorption, or geometrical effects might be needed to explain the line profiles.

However, we cannot say for certain that the radio emission mechanism was the same in January and September, so the possibility remains that we observed a jet-ISM interaction early in the outburst and DT13 observed a baryon-loaded jet later in the outburst. It may be possible to test these interpretations with future observations of that luminous second state. Regardless, it is clear that the observations from 2012 trace a complex relationship between the physics of the radio jet and the X-ray spectrum in 4U 1630--47.

\section{Conclusion}
\label{sec:concl}
The baryon content of astrophysical jets has important consequences for our understanding of their formation and their effects on their environment, but it is difficult to measure directly. \citet{DiazTrigo13} have argued that their detection of relativistically Doppler-shifted emission lines in 4U 1630--47 in September 2012 (MJD 56181) implies the presence of a baryonic jet, most likely launched from the rotating accretion disk (\citealt{BlandfordPayne}). We cannot directly test this interpretation, but in this paper we have used \textit{Chandra} HETGS observations from earlier in the same outburst (January/June, MJD $\sim55954/56081$) to demonstrate that radio emission in 4U 1630--47 is not always associated with relativistic X-ray emission lines. The unique behavior observed with \textit{XMM-Newton} and ATCA later in the outburst may be a special case, dependent on additional processes in the accretion flow around the black hole. However, because much of the observed radio emission from 4U 1630--47 in 2012 is consistent with jet-ISM interactions, we cannot rule out scenarios where the X-ray emission lines originate in a hot outflow that is not physically related to the radio source. Further analysis of the X-ray and radio emission (Neilsen et al., in preparation) may help test our interpretation of this exciting physics.\vspace{-2mm}

\acknowledgements We thank the referee for constructive comments. J.N. acknowledges funding from NASA through the Einstein Postdoctoral Fellowship, grant PF2-130097, awarded by the CXC, which is operated by the SAO for NASA under contract NAS8-03060. GP acknowledges support via an EU Marie Curie Intra-European Fellowship under contract no. FP7-PEOPLE-2012-IEF-331095. We thank N.\ Schulz and D. Huenmoerder for advice regarding plasma models. ATCA is part of the Australia Telescope National Facility which is funded by the Commonwealth of Australia for operation as a National Facility managed by CSIRO.

\label{lastpage}

\end{document}